\documentclass[twocolumn,showpacs,preprintnumbers,prb,fleqn]{revtex4}
\usepackage{graphicx,amsfonts,amsbsy}
\begin{document}

\renewcommand{\Re}{\operatorname{Re}}
\renewcommand{\Im}{\operatorname{Im}}
\newcommand{\Tr}{\operatorname{Tr}}
\newcommand{\sign}{\operatorname{sign}}
\newcommand{\dd}{\text{d}}
\newcommand{\q}{\boldsymbol q}
\newcommand{\p}{\boldsymbol p}
\newcommand{\rr}{\boldsymbol r}
\newcommand{\pp}{p_v}
\newcommand{\vv}{\boldsymbol v}
\newcommand{\I}{{\rm i}}
\newcommand{\pphi}{\boldsymbol \phi}
\newcommand{\ds}{\displaystyle}
\newcommand{\be}{\begin{equation}}
\newcommand{\ee}{\end{equation}}
\newcommand{\bea}{\begin{eqnarray}}
\newcommand{\eea}{\end{eqnarray}}
\newcommand{\Acl}{{\cal A}}
\newcommand{\Rcl}{{\cal R}}
\newcommand{\Tcl}{{\cal T}}
\newcommand{\Tmin}{{T_{\rm min}}}
\newcommand{\Toff}{{\langle \delta T \rangle_{\rm off} }}
\newcommand{\Roff}{{\langle \delta R \rangle_{\rm off} }}
\newcommand{\RoffI}{{\langle \delta R_I \rangle_{\rm off} }}
\newcommand{\RoffII}{{\langle \delta R_{II} \rangle_{\rm off} }}
\newcommand{\dg}{{\langle \delta g \rangle_{\rm off} }}
\newcommand{\rd}{{\rm d}}
\newcommand{\br}{{\bf r}}
\newcommand{\la}{\langle}
\newcommand{\ra}{\rangle}
\newcommand{\nn}{\nonumber}
\newcommand{\da}{\downarrow}
\newcommand{\ua}{\uparrow}

\title{Theory of spin waves in diluted-magnetic-semiconductor quantum wells}
\author{Diego Frustaglia,$^{1,2}$ J\"urgen K\"onig,$^{1,3}$ and Allan H. MacDonald$^4$}
\affiliation{
$^1$Institut f\"ur Theoretische Festk\"orperphysik, Universit\"at Karlsruhe,
76128 Karlsruhe, Germany\\
$^2$NEST-INFM \& Scuola Normale Superiore, 56126 Pisa, Italy\\
$^3$Institut f\"ur Theoretische Physik III, Ruhr-Universit\"at Bochum, 44780 
Bochum, Germany\\
$^4$Department of Physics, University of Texas, Austin,
TX 78712, USA}

\date{\today}

\begin{abstract}

We present a theory of collective spin excitations in 
diluted-magnetic-semiconductor quantum wells in which local magnetic moments 
are coupled via a quasi-two-dimensional gas of electrons or holes.  
In the case of a ferromagnetic state with partly spin-polarized electrons, 
we find that the Goldstone collective mode has 
anomalous $k^4$ dispersion and that for symmetric quantum wells odd parity 
modes do not disperse at all.
We discuss the gap in the collective excitation spectrum which appears when 
spin-orbit interactions are included.  

\end{abstract}


\maketitle


\section{Introduction}

In the emerging field of spin- or magneto-electronics,\cite{P98,WABDvMRCT01} 
the role of the {\it spin} degree-of-freedom in the properties of electronic 
systems is exploited in the design of new functional devices.
The recognition of this additional degree of freedom suggests possibilities
for electrical manipulation beyond the tool-set of conventional electronics
which is based entirely on coupling to the electronic {\it charge}.
The effort to generate and manipulate spin-polarized carriers in a controllable
environment, preferably in semiconductors, has triggered the discovery of 
carrier-induced ferromagnetism \cite{SGFW86,O92} in diluted magnetic 
semiconductors (DMSs).\cite{O98} In these systems 
a few percent of the cations in III-V or II-VI semiconductor compounds are
randomly substituted by magnetic ions, usually Mn, which have local magnetic moments.
The effective coupling between these local moments is mediated by 
free carriers in the host semiconductor compound (holes for p-doped materials 
and electrons for n-doped ones) and can lead to ferromagnetic long-range 
order.  Curie temperatures $T_{\rm c}$ in excess of 100 K have been found in 
bulk (Ga,Mn)As systems.\cite{O98,Gallagher,Samarth} 

\begin{figure} [h]
\begin{center}
\begin{tabular}{cc}
\parbox{4.5cm}{\includegraphics[width=3.5cm,angle=0]{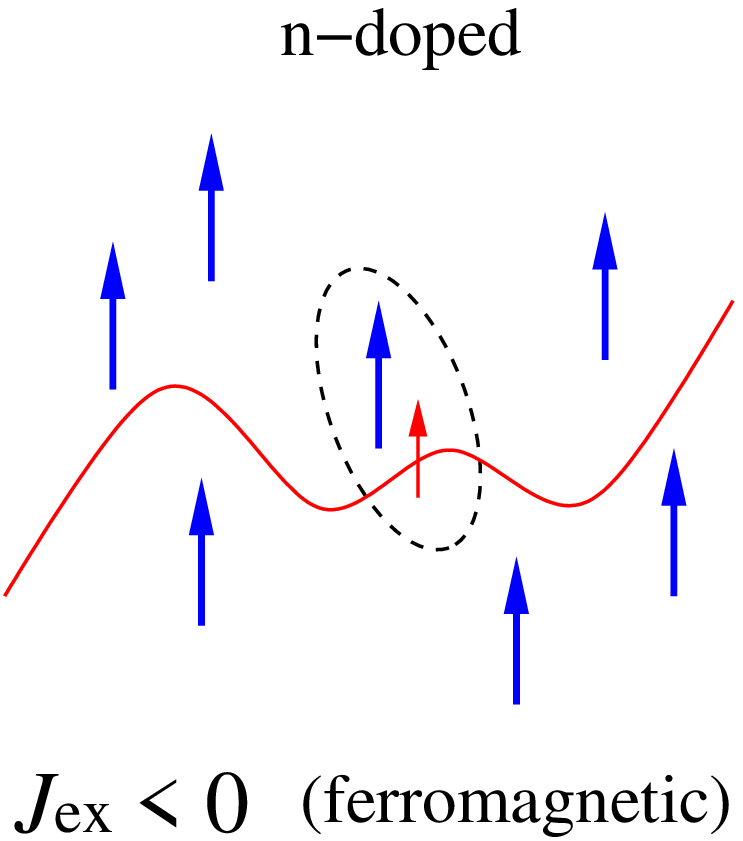}}
&
\parbox{4.5cm}{\includegraphics[width=4cm,angle=0]{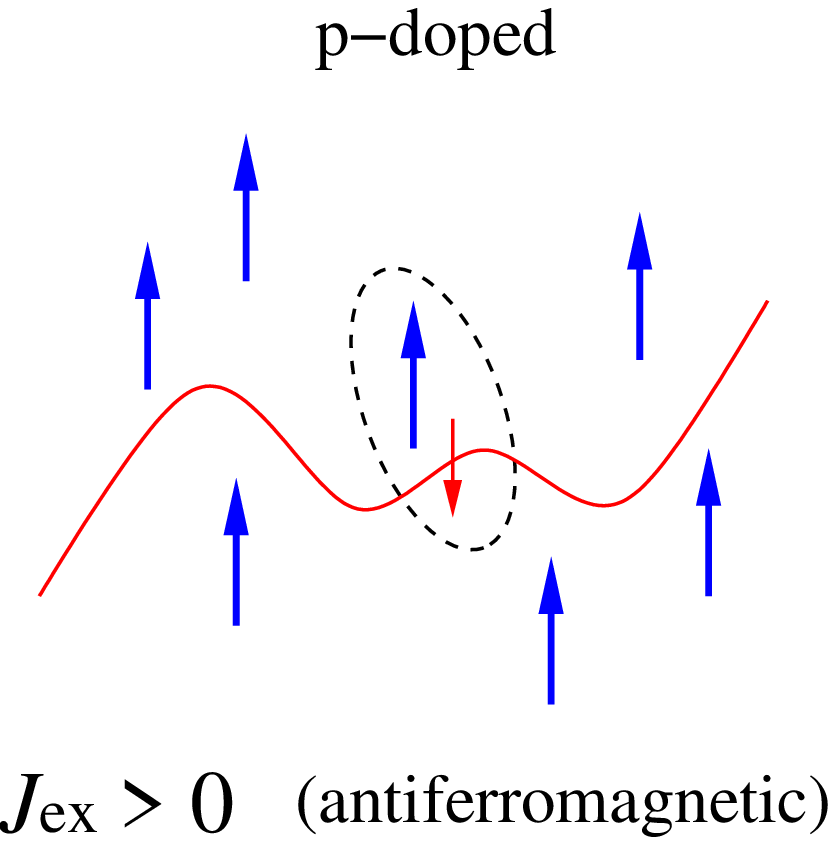}}
\end{tabular}
\end{center}
\caption{
Schematic representation of the exchange coupling between itinerant-carrier
and localized magnetic-impurity spins in n-doped and p-doped DMSs.  When the 
local moments are parallel to each other and the band system is spin-polarized,
the exchange energy can be minimized for either ferromagnetic or antiferromagnetic
interactions.  
}
\label{fig-1}
\end{figure}

One approach to understand the magnetic and optical properties of DMSs is 
based on a phenomenological model of the relevant low-energy degrees of 
freedom.\cite{Furdyna,Dietl94}
In this picture, local $S=5/2$ spins\cite{note6} from Mn$^{2+}$ ions are 
exchange coupled to itinerant carriers of a metallic nature.
In typical samples, the density of free carriers is much smaller than the
Mn ion concentration.
For n-doped materials, the exchange is due to ferromagnetic s-d coupling,
while for p-doped ones it is due to antiferromagnetic p-d coupling, as illustrated 
schematically in Fig.~\ref{fig-1}.
In both cases, the free carriers are believed to mediate an effective
ferromagnetic coupling between the Mn spins, which is typically stronger
than the shorter-range antiferromagnetic direct exchange coupling present in undoped systems.

The reliability of this phenomenological approach has been tested by 
comparing theoretical predictions with experimental findings.
The tendency towards ferromagnetic order and trends in the observed 
$T_{\rm c}$'s, domain structure properties, the anomalous Hall effect, and 
magneto-optical properties, have been successfully described by treating this 
phenomenological model in a mean-field approximation (MFT),\cite{DHdA97,T97,JALMD99,DCFD99,DOMCF00,LJMD00,FRS01,domain,AHE,optics} 
which is analogous to the Weiss mean-field approach for lattice spin models.
In the mean-field theory the local Mn ions are treated as independent but subject to an 
effective magnetic field which originates from their exchange interactions with 
spin-polarized free carriers.
Similarly, the itinerant-carrier system sees an effective field proportional 
to the Mn density and polarization. 
This picture does not account, however, for correlations between
Mn spin configurations and the itinerant carrier state which reduce the energy 
cost of local-moment spin fluctuations that have slow spatial variations. 
As a consequence, MFT systematically overestimates the Curie temperature, 
a problem which is severe for systems with reduced dimensionality,\cite{MW66} 
including the quantum well systems that will be discussed here. 

One prediction that follows from the phenomenological model is that the
system's collective excitations involve correlated dynamics of 
local moment and itinerant spins.  
In the case of bulk DMS systems, we have predicted two branches of collective 
spin waves and discussed their properties\cite{KMd-bulk} as well as their 
impact on limiting the Curie temperature.\cite{Tc}
This analysis of collective excitations requires a theoretical
description beyond MFT, which neglects correlations, and beyond the familiar
Ruderman-Kittel-Kasuya-Yoshida (RKKY) theory of pair-wise carrier-mediated 
interactions, which fails for the systems under
consideration because it assumes a carrier-band spin splitting 
that is small compared to the Fermi energy.  This assumption
is not typically satisfied in doped DMS systems,\cite{HAMSOO98} in part
because the itinerant-carrier concentration is 
usually much smaller than the Mn impurity density.\cite{note10}
Moreover, the RKKY picture also assumes an instantaneous static interaction 
between the magnetic Mn ions, neglecting the retarded character
of the itinerant-carrier response that mediates the interactions. 

An indication\cite{Koenig} that the spin excitations of doped DMS systems 
have collective local-moment and carrier character, even in paramagnetic 
systems, has been provided by recent electron paramagnetic resonance 
experiments \cite{Teran} in n-doped DMS quantum wells.
The aim of the present paper is to extend the previous theoretical work to 
describe the full dispersion of all collective spin excitations in quantum 
wells, their
dependence on the magnetic-ion doping concentration and profile, and on the
free-carrier density.
This is a first step toward the theoretical study of quantum and 
thermal fluctuations in the magnetism of nanostructured 
DMSs which are starting to receive increased attention, partially
because of the possibility of quantum confinement control of 
magnetic properties as in the recent experimental study of a DMS quantum well in 
Ref.~\onlinecite{Ohno-Science}.
Quantum confinement is expected to drastically affect the magnetic properties
of nanostructures.\cite{HWACTDd97,HTSNA97,HTSNSA98,OCMOADOO00,KFACTWdSGBSSWD00,BKBFCTWGD02,NST03,BBFCTWSKWGD03,MKBBFTCG03,WLLDFYWVM03}
In Ref.~\onlinecite{Koenig} only the long-wavelength limit of the lowest
spin-wave branch, the mode that electron paramagnetic resonance probes, was 
considered.

The article is organized as follows. In Sec.~\ref{DT} we develop the 
theoretical tools necessary to address collective excitations in 
doped DMS quantum wells in a general way.
After introducing the many-body quantum Hamiltonian (Sec.~\ref{QWH}) we derive 
an effective action (Sec.~\ref{EA}) that leads to an independent spin-wave theory 
for low temperatures in multi-subband quantum wells (Sec.~\ref{ISWT}). 
Sec.~\ref{ESE} is dedicated to the 
evaluation and discussion of collective spin excitations, 
concentrating on the case in which a single electronic subband is occupied 
and subband mixing is negligible.  We find that odd-parity collective modes of 
doped quantum-well DMS systems are 
dispersionless in this limit, and that the lowest energy Goldstone 
collective mode of ferromagnetic systems 
has anomalous $k^4$ dispersion when the quantum-well
carrier system is not {\it half-metallic} ({\it i.e.} when the carriers are not fully spin-polarized).  Results for dilute and moderate 
Mn doping are shown in Secs.~\ref{ldk} and \ref{hdk}, respectively.  The role 
of spin-orbit coupling, which gives rise to magnetic anisotropy and creates a 
gap in the excitation spectrum of a ferromagnetic system, is discussed in Sec.~\ref{soc}.
A summary and discussion of our results is presented in Sec.~\ref{CC}.

\section{Derivation of the theory}
\label{DT}

\subsection{Hamiltonian}
\label{QWH}

\begin{figure} [h]
\begin{center}
\includegraphics[width=5cm,angle=0]{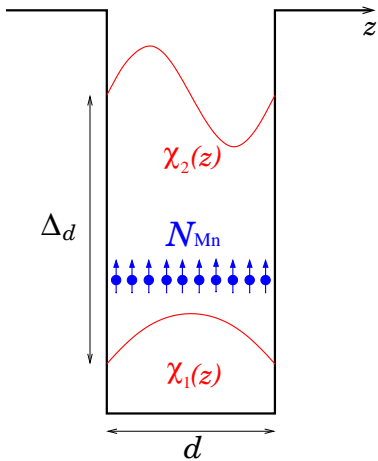}
\end{center}
\caption{
Sketch of a DMS quantum well. Itinerant carriers move 
freely in the $x$-$y$ plane, occupying subbands $\chi_n(z)$ due to quantum 
confinement along the $z$-axis (intersubband energy-gap $\Delta_d$). 
The magnetic-ion doping profile is represented by a continuous Mn-density 
distribution $N_{\rm Mn}(z)$. 
}
\label{fig-2}
\end{figure}

We consider a symmetric quantum well of uniform width $d$ that confines the motion 
of itinerant carriers in the $z$-direction (see Fig.~\ref{fig-2}; we later comment 
the case of asymmetric quantum wells). 
The carriers move freely in the $x$-$y$ plane, occupying one or several 
transverse modes or subbands.
The quantum-well geometry makes it convenient to split the three-dimensional (3D)
spatial coordinate into $({\bf r},z)$, where ${\bf r}$ corresponds to the two-dimensional 
(2D) $x$-$y$-projection.
The field operator for the itinerant carriers $\hat{\Psi}({\bf r},z)$ can be
written as $\hat{\Psi}({\bf r},z)=\sum_{m=1}^M \hat{\psi}_m({\bf r})\chi_m(z)$,
where $m=1,\ldots,M$ labels the subband number, 
$\chi_m(z)=\sqrt{2/d}~\sin(m \pi z/d)$ is the real
wave function for subband $m$, which satisfy the orthonormality condition
$\int_0^d {\rm d}z~\chi_m(z) \chi_{m'}(z)=\delta_{m,m'}$, and
$\hat{\psi}_m({\bf r})$ is a spinor with components 
$\hat{\psi}_{\sigma,m}({\bf r})$.  As indicated above we will adopt
particle-in-a-box wavefunctions for explicit calculations, although
this approximation plays no critical role in our theory and can 
easily be relaxed.  The transverse wave function $\chi_m(z)$ 
degree-of-freedom will later be taken to be frozen in its ground state; this 
is normally a good approximation except in wide quantum wells.
The in-plane degrees of freedom are described by  
{\it fluctuating} spinor fields $\hat{\psi}_{\sigma,m}({\bf r})$. 
The magnetic impurities are randomly distributed within the quantum well.

The fact that the Mn density in typical quantum well systems is very
much larger than the carrier density suggests the replacement of the random distribution of 
local Mn magnetic moments by a continuous density $N_{\rm Mn}(z)$, thereby neglecting 
disorder in the Mn ion locations.\cite{note7}  This leaves us with a situation in 
which a growth direction degree of freedom exists for the local moment spins, but not for
the quantum well electrons.  It is the quasi-3D character of local moments that are 
coupled together by quasi-2D electrons that is responsible for unusual aspects of the
collective excitation spectrum that we will discuss later.
The Debye-like continuum approximation we use
for the Mn ion density distribution will, of course, fail for modes
that involve either in-plane or growth-direction spatial variation on a scale
shorter than the distance between Mn ions, as we discuss later.
The dependence of $N_{\rm Mn}(z)$ on the growth-direction coordinate $z$
allows for the possibility of a non-uniform doping profile in the quantum well.
The two-component spinors we use for the quantum-well electron fields restrict our 
attention to circumstances in which the electric subbands occur in pairs with 
identical orbital wavefunctions, {\it i.e.} to hole quantum-well subbands with 
small heavy-light mixing or to electron subbands.   Generalizing to
arbitrarily spin-orbit coupled systems considerably complicates the notation we use below 
and, in the case of an external field, complicates the theory considerably because of 
the interplay between orbital and Zeeman coupling.
We will for the most part restrict our attention to n-doped quantum wells.  

The total Hamiltonian $H$ consist of four terms: 
$H=H_{\rm kin}+H_{\rm Z}+H_{\rm ex}+H_{\rm D}$. In the presence of a magnetic 
field ${\bf B}=\nabla \times {\bf A}$, the kinetic term for 
carriers of charge $e$ reads
\bea
H_{\rm kin}&=&\int {\rm d}^2r \int_0^d {\rm d}z \times 
\label{Hkin} \\
&~&\hat{\Psi}^\dag({\bf r},z) 
\left[ \frac{-\hbar^2}{2 m^*} \tilde{\nabla}^2 + V(z) - \mu \right] 
\hat{\Psi}({\bf r},z) \nn \\ 
&=&\int {\rm d}^2r \sum_{m=1}^M \sum_\sigma 
\hat{\psi}_{\sigma,m}^\dag({\bf r}) \left[ \frac{-\hbar^2}{2 m^*} 
\tilde{\nabla}_{\bf r}^2 - \mu_m' \right] \hat{\psi}_{\sigma,m}({\bf r}), \nn
\eea
where $\tilde{\nabla}_{({\bf r})}=\nabla_{({\bf r})}-(i e/\hbar c) {\bf A}$,
$V(z)$ is the quantum-well confining potential, and $\mu'_m=\mu-\epsilon_m$ is the
effective subband-dependent chemical potential of the quasi-2D carrier gas, 
with $\epsilon_m$ the subband quantization energy. 
Here, we assumed a parabolic dispersion for the free carriers, with an 
effective mass $m^*$.
This is well justified for n-doped systems, which have s-band conduction 
electrons.
For hole doped systems, which have p-band valence carriers, this approximation 
is often useful for qualitative discussions.  The Zeeman term is 
\bea
H_{\rm Z}= 
\mu_{\rm B} {\bf B} \cdot \int {\rm d}^2r \int_0^d {\rm d}z
\left[ g_e \hat{\bf s}({\bf r},z)+g_{\rm Mn} {\bf S}({\bf r},z)\right]
\label{Hz},
\eea
where $\mu_{\rm B}>0$ is the Bohr magneton,
\bea
\label{s}
\hat{\bf s}({\bf r},z) &=& \frac {1}{2} \hat{\Psi}^\dag({\bf r},z)~\boldsymbol{\tau}~
\hat{\Psi}({\bf r},z) \\
&=&\sum_{m,m'} \sum_{\sigma,\sigma'} 
\chi_m(z)\chi_{m'}(z)~
\hat{\psi}_{\sigma,m}^\dag({\bf r})
\frac{\boldsymbol{\tau}_{\sigma \sigma'}}{2}
\hat{\psi}_{\sigma',m'}({\bf r}) \nn
\eea
is the quantum-well carrier spin density (with Pauli matrix vector 
$\boldsymbol{\tau}$), and ${\bf S}({\bf r},z)$ is the spin density of the Mn 
subsystem. The coupling between the carrier spins and the local Mn spins 
is described by
\bea
H_{\rm ex}&=&J_{\rm ex}\int {\rm d}^2r \int_0^d {\rm d}z~{\bf S}({\bf r},z)
\cdot \hat{\bf s}({\bf r},z),
\label{Hex}
\eea
where $J_{\rm ex}<0$ corresponds to ferromagnetic and $J_{\rm ex}>0$ 
to antiferromagnetic coupling ({\it i.e.} to n- and p-doped host semiconductors, 
respectively). In symmetric quantum wells spin-orbit interactions are described
by the Dresselhaus Hamiltonian\cite{note9}
\bea
H_{\rm D}&=&\gamma \int {\rm d}^2r \int_0^d {\rm d}z~\hat{\Psi}^\dag({\bf r},z)
~k_z^2 \left(- \tau_x k_x + \tau_y k_y \right) 
\hat{\Psi}({\bf r},z) \nn \\
&=&\gamma \int {\rm d}^2r 
\sum_{m,m'=1}^M \sum_{\sigma,\sigma'} 
~\langle k_z^2 \rangle_{m,m'} \times
\nn \\
&~&\hat{\psi}_{\sigma,m}^\dag({\bf r}) \left(- \tau_x k_x + \tau_y k_y \right)_{\sigma \sigma'} \hat{\psi}_{\sigma',m'}({\bf r}),
\label{HD}
\eea
where $\langle k_z^2 \rangle_{m,m'}=\int_0^d {\rm d}z~\chi_m(z)~k_z^2~\chi_{m'}(z)=\langle k_m^2 \rangle \delta_{m,m'}$ with $\langle k_m^2 \rangle=(m \pi/d)^2$ when particle-in-a-box orbitals are 
used. 
The above spin-orbit Hamiltonian $H_{\rm D}$ leads to a $z$-oriented 
magnetic easy-axis as we have shown in earlier work\cite{Koenig} and discuss 
later. For comments on spin-orbit coupling in the case of asymmetric quantum wells 
see Sec. \ref{soc}.

\subsection{Effective action}
\label{EA}

In analogy to our earlier work\cite{KMd-bulk} on bulk DMS ferromagnets, we
want to describe elementary spin excitations in the DMS quantum well in a 
language where the itinerant-carrier degrees of freedom are integrated out.
This leads to a retarded free-carrier mediated interaction between the
Mn-ion $S=5/2$ spins.
We are interested in small spin fluctuations about the mean-field magnetic
state.  The ground state of experimental doped quantum well DMS systems 
have sometimes been found to be ferromagnetic,\cite{Ohno-Science,HWACTDd97,HTSNA97,HTSNSA98,OCMOADOO00,KFACTWdSGBSSWD00,BKBFCTWGD02,NST03,BBFCTWSKWGD03,MKBBFTCG03,WLLDFYWVM03} 
and sometimes exhibit complex spin-glass behavior. 
It is quite possible that the complex spin-glass states that
sometimes occur are due to disorder effects that are not essential and can 
in principle be avoided, due for example to inhomogeneities in the 
Mn ion distribution, substitutional Mn ions, or other defects.  
In any event, the theory we discuss assumes a mean-field state in 
which all Mn ions are aligned.  When this simple state is not the ground
state of the system, or when we want to describe the collective excitations 
of a system that is above its ferromagnetic transition temperature, our 
theory will apply only in an external magnetic field that is strong
enough to achieve substantial Mn ion spin polarization.   
We choose this field to be oriented in the $\hat z$ direction: ${\bf B}=(0,0,B)$.
Because $g_{\rm Mn}>0$, the Mn spins then tend to align along the opposite  
direction.  In the case of ferromagnets with anisotropy, we choose the $\hat z$ direction to be
along an easy-axis.  

It is convenient to represent the $S=5/2$ spins by 
Holstein-Primakoff (HP)\cite{HP40} bosons.
For small fluctuations around the mean-field state the spin density
${\bf S}({\bf r},z)$ is approximated by $S^+ \approx \bar{\omega} 
\sqrt{2 N_{\rm Mn}(z) S}$, $S^- \approx \omega \sqrt{2 N_{\rm Mn}(z) S}$, 
and $S^z = \bar{\omega}\omega - N_{\rm Mn}(z) S$, 
where the complex variables $\bar{\omega}$, $\omega$ are boson creation and 
annihilation operators that become bosonic coherent
state labels in the path-integral formalism we employ.
The vacuum with no HP bosons corresponds to full (negative) polarization of 
the Mn system,
while the creation of a HP boson describes an increase in the total Mn spin by one unit.
The partition function $Z$ of the compound system is calculated using 
a coherent-state path integral representation
\bea
Z=\int {\cal D}(\bar{\psi} \psi){\cal D}(\bar{\omega}\omega)~ 
e^{-\int_0^\beta {\rm d}\tau~L(\bar{\psi} \psi,\bar{\omega}\omega)}
\eea
with ${\cal D}(\bar{\psi} \psi) \equiv {\cal D}(\bar{\psi}_1 \psi_1) \ldots {\cal D}(\bar{\psi}_M \psi_M)$ and the Lagrangian
\begin{widetext}
\bea
L(\bar{\psi} \psi,\bar{\omega}\omega)&=& 
\int {\rm d}^2r 
\left[\sum_{m=1}^M \bar{\psi}_m({\bf r,\tau})
\partial_\tau \psi_{m}({\bf r,\tau}) 
+ \int_0^d {\rm d}z~\bar{\omega}({\bf r},z,\tau) 
\partial_\tau \omega({\bf r},z,\tau) \right] 
+ H(\bar{\psi} \psi,\bar{\omega}\omega),  
\eea
\end{widetext}
where the Grassmann numbers $\bar{\psi}=(\bar{\psi}_1,\ldots,\bar{\psi}_M)$ and
$\psi=(\psi_1,\ldots,\psi_M)$ describe fermion (itinerant carrier) fluctuations 
within each of the $M$ subbands. 

Since the Hamiltonian is bilinear in the fermionic fields, we can integrate 
them out and arrive at a representation for the
bosonic partition function of the form $Z=\int {\cal D}(\bar{\omega} \omega) 
\exp(-S_{\rm eff}[\bar{\omega} \omega])$, with an effective action 
\begin{widetext}
\bea
\label{Seff}
S_{\rm eff}[\bar{\omega} \omega]=\int_0^\beta {\rm d}\tau \int {\rm d}^2 r 
\int_0^d {\rm d}z~\left[ \bar{\omega}({\bf r},z,\tau) 
\partial_\tau \omega({\bf r},z,\tau) + g_{\rm Mn} \mu_{\rm B} {\bf B} \cdot {\bf S}(\bar{\omega} \omega) \right] -\ln[\det G^{-1}(\bar{\omega} \omega)].
\eea
\end{widetext}
The total kernel $G^{-1}(\bar{\omega} \omega)$ can be split into a mean-field
part, which does not depend on the bosonic (Mn spin excitation) fields 
$\bar{\omega}$ and $\omega$, and a fluctuating 
part by writing $G^{-1}(\bar{\omega} \omega)=G_{\rm MF}^{-1} + 
\delta G^{-1}(\bar{\omega} \omega)$, with
\bea
\label{Gmf}
\left(G_{\rm MF}^{-1}\right)_{m,m'}&=&
\left[ \partial_\tau-\frac{\hbar^2}{2 m^*} 
\tilde{\nabla}_{\bf r}^2 - \mu'_m \right] \delta_{m,m'} \\
&+&\frac{1}{2} \left(g_e \mu_{\rm B} B \delta_{m,m'} - \Delta_{m,m'}\right) 
\tau_z \nn \\
&+& \gamma \langle k_m^2 \rangle \left(-\tau_x k_x + \tau_y k_y \right)
\delta_{m,m'}, \nn \\
\label{dG}
\delta G_{m,m'}^{-1}(\bar{\omega} \omega)&=& \frac {J_{\rm ex}}{2} \int_0^d {\rm d}z~\chi_m(z) \chi_{m'}(z) \times\\
&&\left[\sqrt{2 N_{\rm Mn}(z) S}~(\bar{\omega}~\tau^- +\omega~\tau^+)+\bar{\omega} \omega \tau_z \right]. \nn
\eea
The exchange coupling contributes to the conduction band spin-splitting in 
$G_{\rm MF}$ through the mean-field interaction
$\Delta_{m,m'}=J_{\rm ex} (\bar{N}_{\rm Mn})_{m,m'} S$, where 
$(\bar{N}_{\rm Mn})_{m,m'}=\int_0^d {\rm d}z~\chi_m(z) \chi_{m'}(z) N_{\rm Mn}
(z)$.\cite{note1}
We recognize here that coupling to a Mn-spin system with an inhomogeneous 
doping profile $N_{\rm Mn}(z)$ leads to mixing of the quantum-well subbands.
These intersubband interactions are present at the mean-field level, as seen in Eq.~(\ref{Gmf}), but
also appear in the term which expresses the coupling between carriers and 
local moment fluctuations, Eq.~(\ref{dG}). 

\subsection{Quantum-well subband decoupling and independent spin-wave theory}
\label{ISWT}

The above picture simplifies considerably when quantum-well subband mixing
is negligible,\cite{LeePRB} {\it i.e.} when the energy gap $\Delta_d$ (see Fig.~\ref{fig-2}) is much larger than 
the spin-splitting energies $\Delta_{m,m'}$. This regime can be reached either by
narrowing the quantum well ($\Delta_d \rightarrow \infty$), or by 
diluting the Mn doping ($\Delta_{m,m'} \rightarrow 0$),
independently of the number of occupied subbands $M$, which is 
controlled by the carrier density.  In this limit we arrive at a Green's function that is diagonal in subband space, 
with $G^{-1}_m=(G_{\rm MF}^{-1})_m+\delta G^{-1}_m$,
\bea
\label{Gmfm}
(G_{\rm MF}^{-1})_m&=&\left[ \partial_\tau-\frac{\hbar^2}{2 m^*} 
\tilde{\nabla}_{\bf r}^2 - \mu'_m\right] \\
&+& \frac {1}{2} (g_e \mu_{\rm B} B -\Delta_m) \tau_z \nn \\
&+& \gamma \langle k_m^2 \rangle \left(-\tau_x k_x + \tau_y k_y \right), \nn \\
\label{dGm}
\delta G^{-1}_m(\bar{\omega} \omega)&=& \frac {J_{\rm ex}}{2} \int_0^d {\rm d}z~\chi^2_m(z)\times\\
&~&\left[ \sqrt{2 N_{\rm Mn}(z) S}~(\bar{\omega}~\tau^- +\omega~\tau^+)+\bar{\omega} \omega \tau_z \right] \nn.
\eea
Here $\Delta_m=\Delta_{m,m}$ is the subband-dependent exchange contribution to
the itinerant carrier mean-field splitting.  This approximation leads to a convenient subband separation
$\ln[\det G^{-1}]=\sum_{m=1}^M \ln[\det G^{-1}_m]$ in Eq.~(\ref{Seff}).

Expanding $\ln[\det G^{-1}_m]$ up to second order in 
$\delta G^{-1}_m$,  $\ln[\det G^{-1}_m] = 
{\rm tr}[\ln (G_{\rm MF}^{-1})_m] + 
{\rm tr}[(G_{\rm MF})_m \delta G^{-1}_m]-
(1/2) {\rm tr}[(G_{\rm MF})_m \delta G^{-1}_m (G_{\rm MF})_m 
\delta G^{-1}_m] + \ldots $, and collecting all contributions up to quadratic order in 
$\bar{\omega}$ and $\omega$ we arrive at an independent 
spin-wave theory where the spin excitations are treated as non-interacting
HP bosons. This is a good approximation for temperatures well below the 
maximum of the ferromagnetic transition temperature and/or the temperature 
defined by Zeeman coupling to the external field, in which case 
spin excitation amplitudes are small.
Fourier transforming the resulting spin-wave action (keeping 
$z$ in real space and defining bosonic Matsubara frequencies $\nu_n$) 
we obtain
\bea
S_{\rm eff}[\bar{\omega}\omega]&=&\frac{1}{\beta} 
\sum_n \int \frac{{\rm d}^2k}{(2 \pi)^2} \int \frac{{\rm d}^2k'}{(2 \pi)^2} 
\int_0^d {\rm d}z \int_0^d {\rm d}z'~\times \nn \\
&&\bar{\omega}({\bf k},z,\nu_n) D^{-1}({\bf k},{\bf k'},z,z',\nu_n) 
\omega({\bf k'},z',\nu_n).~~~~~
\label{SeffD}
\eea 
The kernel of the quadratic action (\ref{SeffD}) is the inverse of the 
spin-wave propagator $D({\bf k},{\bf k'},z,z',\nu_n)$ and is given by
\begin{widetext}
\bea
D^{-1}({\bf k},{\bf k'},z,z',\nu_n)&=&\left[-i \nu_n+g_{\rm Mn} \mu_B B - 
\frac 
{J_{\rm ex}}{2} \sum_{m=1}^M (n^\da_m-n^\ua_m) \chi_m^2(z)\right]
\delta(z-z') \delta({\bf k}-{\bf k'})
\nn \\
&+&\frac {J_{\rm ex}^2}{2} S~\sqrt{N_{\rm Mn}(z)N_{\rm Mn}(z')}~\sum_{m=1}^M 
\chi_m^2(z) \chi_m^2(z') \sum_{\alpha,\alpha'}~ 
\frac {f(\epsilon_{m,\alpha}^\da)
-f(\epsilon_{m,\alpha'}^\ua)}{i \nu_n + \epsilon_{m,\alpha}^\da - 
\epsilon_{m,\alpha'}^\ua}~\Phi_m^{\alpha,\alpha' *}({\bf k})~ 
\Phi_m^{\alpha,\alpha'}({\bf k'}).
\label{D}
\eea
\end{widetext}
We arrive at Eq.~(\ref{D}) by introducing a wave-function representation of 
the mean-field Green's function of Eq.~(\ref{Gmfm}) 
\bea
(G_{\rm MF})_m^\sigma({\bf r'},{\bf r}, \nu_n) =\sum_\alpha \frac
{\phi_{m,\alpha}^\sigma({\bf r'}) \phi_{m,\alpha}^{\sigma *}({\bf r})}
{i \nu_n-\epsilon_{m,\alpha}^\sigma}, 
\label{GMFr}
\eea 
where $\phi_{m,\alpha}^\sigma({\bf r})$ are the 2D mean-field itinerant 
carrier eigenstates for spin $\sigma$ 
and subband $m$, with energy $\epsilon_{m,\alpha}^\sigma=
\epsilon_\alpha+(\sigma/2)(g_e \mu_{\rm B} B-\Delta_m) \sqrt{1+ 4 \epsilon^{{\rm so}}_{m,\alpha}/(g_e \mu_{\rm B} B-\Delta_m)}-\mu'_m$. \cite{sakurai} The index 
$\alpha$ accounts for quantum-numbers associated with orbital motion,
with kinetic and spin-orbit energies $\epsilon_\alpha$ and  
$\epsilon^{{\rm so}}_{m,\alpha}=|\langle \downarrow,m,\alpha|\gamma k_z^2(-\tau_x k_x + \tau_y k_y)|\uparrow,m,\alpha \rangle|^2/(g_e \mu_{\rm B} B-\Delta_m)$, respectively.
To first order in spin-orbit coupling
($\epsilon^{{\rm so}}_{m,\alpha} \ll (g_e \mu_{\rm B} B-\Delta_m)$), 
the eigenenergy is  
$\epsilon_{m,\alpha}^\sigma = \epsilon_\alpha+(\sigma/2) (g_e \mu_{\rm B} B-\Delta_m)+\sigma \epsilon^{{\rm so}}_{m,\alpha}-\mu'_m$. 
In Eq.~(\ref{D}) $f(\epsilon)$ is the Fermi distribution, $n^\sigma_m$ is the 2D 
mean-field itinerant carrier spin density for the $m$,$\sigma$ subband obtained by summing over
occupied states, and we have introduced Fourier-transform factors defined by 
\bea
\Phi_m^{\alpha,\alpha'}({\bf k})=\int {\rm d}^2r~\exp(i {\bf k}\cdot {\bf r})~ 
\phi_{m,\alpha}^\ua({\bf r}) \phi_{m,\alpha'}^{\da *}({\bf r}).
\label{Phi}
\eea

The first line on the r.h.s. of Eq.~(\ref{D}) is local in space.
It represents the mean-field expression for the exchange field that the Mn 
spins experience.
The second line is nonlocal in space and describes correlation effects 
that occur because of the (space- and $m$-subband-dependent) response of the 
quantum-well carriers to Mn spin orientations. We note that $D^{-1}$ is
not a function of $z-z'$ only, because of the absence of translational
symmetry in the $z$-direction.  Additionally, a 2D Debye cutoff
$k_{\rm D}^2=4 \pi N_{\rm Mn}/(N/d)$ ensures that our continuum approximation
has the correct number of magnetic-impurity degrees of freedom. 
[Here, $N_{\rm Mn}=\int_0^d {\rm d}z N_{\rm Mn}(z)/d$~~\cite{note3} and
$N$ is the number of growth-direction modes included in the theory as we explain below.
We associate $N$ with the mean number of Mn ions encountered on crossing the 
quantum well; {\it e.g.} for isotropic doping $N \sim N_{\rm Mn}^{1/3} d$.]

We comment now on the factors $\Phi_m^{\alpha,\alpha'}({\bf k})$ which are
trivial for the plane-wave functions of field-free systems.  They are included
to allow us to simply account for the consequences of orbital coupling
of itinerant carriers to magnetic fields. 
The index $\alpha$ includes both  
the Landau levels index and the gauge-dependent index for states within 
a Landau level.  At zero field the mean-field eigenstates 
$\phi_{m,\alpha}^\sigma({\bf r})$ are plane waves with momentum 
$\alpha \equiv {\bf q}$, kinetic energy 
$\epsilon_{\bf q}=\hbar^2 q^2/(2 m^*)$ and spin-orbit energy 
$\epsilon^{\rm so}_{m,{\bf q}}=\gamma^2 \langle k^2_m \rangle^2 q^2/(g_e \mu_{\rm B} B-\Delta_m)$. 
In either case Eq.~(\ref{D}) is diagonal in the in-plane momentum ${\bf k}$ and we 
denote its diagonal
elements by $D^{-1}({\bf k},z,z',i \nu_n)$.  The case of a uniform magnetic
field was considered in Ref.~\onlinecite{Koenig} and will not be discussed further in this paper. 

\section{Elementary spin excitations}
\label{ESE}

Even the bulk-like epitaxially grown thin film samples studied in typical experiments 
do not contain a very large number of occupied 2D carrier subbands. Our formalism 
could in principle be used to calculate the collective modes of thin films, 
taking account of the variation in Mn density across the film, although the approach 
becomes numerically cumbersome when more than a few subbands are occupied.
It is likely, though, that greater insight into vertical inhomogeneity effects 
in thin-films can be obtained with more approximate approaches.\cite{RRLZFJ03}
We limit the discussion here to true quantum-well samples in which a single
subband is occupied ($M=1$) and subband mixing can be neglected.  For definiteness we 
concentrate on the case of constant Mn density $N_{\rm Mn}(z)=N_{\rm Mn}$. 
Similarly we assume that the state we are studying is an ordered ferromagnetic phase 
with the $\hat z$-direction magnetic easy-axis that is favored by (weak) spin-orbit coupling.
The situation in which the magnetization direction has been
reoriented by an external magnetic field is readily included in the formalism as explained above.
Generalizations to the cases of multiple subbands, and  
inhomogeneous Mn density are straightforward.

Collective spin excitation dispersion $\Omega({\bf k})$
branches are located by finding the frequencies at which 
the determinant, $\det [D^{-1}({\bf k},z,z',i \nu_n=\Omega)]$, 
of the quadratic action kernel in Eq.~(\ref{SeffD}) vanishes. 
The continuum of spin-flip particle-hole (Stoner) excitations is 
located by identifying the ${\bf k}$-dependent frequency range over
which ${\rm Im} [D^{-1}({\bf k},z,z',i \nu_n)]$ is non-zero 
after the analytic continuation $i \nu_n \rightarrow \Omega+i 0^+$.   

\subsection{Stoner continuum}

We start by evaluating the continuum of Stoner 
excitations introduced above. They correspond to flipping a single spin in 
the itinerant carrier subsystem and typically have relatively large energies 
of the order of the itinerant carrier mean-field spin splitting 
$\Delta= J_{\rm ex} N_{\rm Mn}S$. The continuum is obtained by 
determining the conditions 
for ${\rm Im} [D^{-1}({\bf k},z,z',\Omega+i 0^+)] \neq 0$.
With this aim it is 
convenient to define the dimensionless carrier spin-polarization 
$p=(n^\ua-n^\da)/(n^\ua+n^\da)$ and the Fermi energy of the majority-spin 
carrier band $\epsilon_{\rm s}=\mu'+|\Delta|/2$, where $\mu'$ is the effective chemical 
potential of the 2D carrier gas. For half-metallic carriers 
($|p|=1$, $\epsilon_{\rm s} \le |\Delta|$, see Appendix \ref{AppSP}) these 
excitations carry spin $S_z=\pm 1$, depending on the sign of $\Delta$ ({\it i.e.} on 
whether the coupling between carriers and Mn ions is ferromagnetic or antiferromagnetic). 
On the other hand, for partly polarized carriers ($|p| < 1$, 
$\epsilon_{\rm s} > |\Delta|$, see Appendix \ref{AppSP}) excitations carrying 
both 
$S_z= 1$ and $S_z=- 1$ contribute, independent of the sign of $\Delta$. 
In the absence of spin-orbit coupling ($\gamma=0$), 
one finds a continuum of excitations with dispersion lying between 
the curves 
$-\Delta - {\rm sign}[\Delta] \epsilon_{\bf k} \pm 2 \sqrt{\epsilon_{\rm s} 
\epsilon_{\bf k}}$ for $\epsilon_{\rm s} \le |\Delta|$, and also between 
$-\Delta + {\rm sign}[\Delta] \epsilon_{\bf k} \pm 2 
\sqrt{(\epsilon_{\rm s}-|\Delta|) \epsilon_{\bf k}}$ for 
$\epsilon_{\rm s} > |\Delta|$. For small spin-orbit coupling, the energy width 
of the Stoner continuum does not vanish at ${\bf k}=0$, instead approaching 
the width  
\bea
\Delta \Omega = 
4 \frac {\gamma^2 \langle k_z^2 \rangle^2 m^* \epsilon_{\rm s}}{\hbar^2 |\Delta|}.
\eea
In the case of multi-subband quantum wells ($M > 1$) 
multiple continua arise, each of them associated with the corresponding 
subband by means of $\Delta_m$, $\langle k_m^2 \rangle$ and $\mu'_m$.

\subsection{Spin-wave modes}
\label{DMnD}

The number of collective modes that appear in our theory depends on the 
doping concentration and the width of the quantum well.
It is natural to choose the mean number $N$ of Mn ions along the 
$z$-direction as a dimensional cutoff for the representation of the inverse
propagator $D^{-1}({\bf k},z,z',\Omega)$.
This motivates the choice of an appropriate basis of $N$ orthonormal 
excitation profiles $\{ \omega_n(z)\}$, with $0 \le n \le N-1$ and
$\int_0^d {\rm d}z~\omega_n(z) \omega_{n'}(z)=\delta_{nn'}$, for expanding 
Eq.~(\ref{D}), {\it i.e.},
\bea
D^{-1}_{nn'}({\bf k},\Omega)=\int\limits_0^d {\rm d}z \int\limits_0^d 
{\rm d}z'
\omega_n(z) D^{-1}({\bf k},z,z',\Omega) \omega_{n'}(z'). \,\,
\label{Dnm}
\eea
We later solve $\det[D^{-1}_{nn'}({\bf k},\Omega)]_{N \times N}=0$ for 
$\Omega$ and obtain a set of $N+1$ solutions $\{\Omega_{(l)}({\bf k})\}$ 
with the mode profiles $\omega_{(l)}(z)=\sum_n c^{(l)}_n \omega_n(z)$.
The coefficients ${\bf c}^{(l)}=(c^{(l)}_0,c^{(l)}_1,\ldots,c^{(l)}_{N-1})$
are obtained from 
$[D^{-1}_{nn'}({\bf k},\Omega_{(l)})]_{N \times N} {\bf c}^{(l)}=0$.
We combine this procedure with a Debye cut-off of the 2D wavevectors
to get the correct number of magnetic degrees of freedom.  This approximate 
procedure, a silent partner of the continuum Mn density approximation that we 
use to avoid dealing with disorder, obviously breaks down to some degree for
the shortest wavelength modes which must be sensitive to the discreteness of 
the magnetic degrees of freedom.  The procedure should be accurate for 
longer wavelength modes, however, and we believe that it gives a good qualitative
description of the overall spectrum.  We employ it without further comment in the 
rest of the paper.  

We assume that the magnetization direction of the Mn spins located at the 
borders of the quantum well is not fixed by an anisotropy field or magnetic
coupling to an adjacent layer.
Then we can use free-end boundary conditions for the magnetic excitations, 
${\rm d}\omega_n(z)/{\rm d}z=0$ at $z=0,d$.
This determines the choice of the basis functions
\bea
\omega_n(z)=a_n \cos\left(\frac{n \pi z}{d}\right),~~~a_n=\left\{ 
\begin{array}{cc}
\sqrt{1/d} & {\rm for}~n=0 \\
\sqrt{2/d} & {\rm for}~n \ge 1
\end{array} \right..
\label{wn}
\eea  

We now calculate the matrix elements $D^{-1}_{n n'}({\bf k},\Omega)$ using
Eqs.~(\ref{D})-(\ref{wn}) for $M=1$ and constant $N_{\rm Mn}$ in the 
absence of an external magnetic field. 
The quantum-well subband is defined by the wave function 
$\chi(z)=\sqrt{2/d}~\sin(\pi z/d)$. 
We find
\bea
D^{-1}_{n n'}({\bf k},\Omega) = -\Omega \delta_{nn'} 
+ x_{\rm s} |\Delta| \left[M_1 + I({\bf k},\Omega) M_2 \right]_{nn'}, 
\label{Dnm-2}
\eea
with the $N\times N$-matrices
\bea
M_1&=&\left( \begin{array}{cccccc}
1 & 0 & -1/\sqrt{2} & 0 & 0 & \ldots\\
0 & 1/2 & 0 & -1/2 & 0 & \ldots\\
-1/\sqrt{2} & 0 & 1 & 0 & -1/2 & \ldots \\
0 & -1/2 & 0 & 1 & 0 & \ldots \\
0 & 0 & -1/2 & 0 & 1 & \ldots \\
\vdots & \vdots & \vdots & \vdots & \vdots & \ddots
\end{array} \right)_{N \times N} \label{M2}\\
\nn \\
M_2&=&\left( \begin{array}{ccccc}
1 & 0 & -1/\sqrt{2} & 0 & \ldots\\
0 & 0 & 0 & 0 & \ldots\\
-1/\sqrt{2} & 0 & 1/2 & 0 & \ldots \\
0 & 0 & 0 & 0 & \ldots \\
\vdots & \vdots & \vdots & \vdots & \ddots
\end{array} \right)_{N \times N}.
\label{M3}
\eea
In Eq.~(\ref{Dnm-2}) we have defined the ratio of the free-carrier spin-density 
to the Mn spin density
$x_{\rm s}=|n^\ua-n^\da|/(2 N_{\rm Mn} S d)$, which typically satisfies $x_{\rm s} \ll 1$, and the 
dimensionless integral
\bea
I({\bf k},\Omega) =
\frac {|\Delta|}{|n^\ua-n^\da|} \int \frac {{\rm d}^2q}{(2 \pi)^2}~ 
\frac {f(\epsilon_{\bf q}^\da)
-f(\epsilon_{{\bf q}+{\bf k}}^\ua)}{\Omega + \epsilon_{\bf q}^\da - 
\epsilon_{{\bf q}+{\bf k}}^\ua}.
\label{IK}
\eea
The last can be evaluated analytically for $T=0$ in the absence of spin-orbit coupling,
as described briefly in Appendices \ref{AppSP} and \ref{AppCC}.
 
The term containing $M_1$ corresponds to the term proportional to $J_{\rm ex}$ in 
Eq.~(\ref{D}) and describes the mean-field exchange interaction between Mn spins
with free-carrier spins.
The appearance of off-diagonal matrix elements, indicating a mixing of basis
functions for the Mn spin excitations, is of geometric origin, determined by 
the projection $\int_0^d {\rm d}z~\chi^2(z) \omega_n(z) \omega_{n'}(z)$.
The nonlocal correlations are accounted for by the term containing $M_2$, 
which corresponds to the term proportional to $J_{\rm ex}^2$ in Eq.~(\ref{D}).
Mixing appears here also, determined this time by
$[\int_0^d {\rm d}z~\chi^2(z) \omega_n(z)][\int_0^d {\rm d}z~\chi^2(z) 
\omega_{n'}(z)]$.
The structure of the matrices in Eqs.~(\ref{M2}) and (\ref{M3}) shows that
basis functions with different parity $\{\omega_n(z)\}$ do not mix.\cite{note4}
This allows us to write the  
expanded kernel (\ref{Dnm-2}) as the (external) product of two matrices 
corresponding to even ($+$) and odd ($-$) modes: 
$[D^{-1}_{n n'}({\bf k},\Omega)]_{N \times N}= 
[D^{-1}_{n n'}({\bf k},\Omega)]^+_{N^+ \times N^+} \otimes 
[D^{-1}_{n n'}({\bf k},\Omega)]^-_{N^- \times N^-}$, with $N^+ + N^-=N$. 
Spin modes obtained as solutions of 
$\det[D^{-1}_{nn'}({\bf k},\Omega)]_{N \times N}=0$ can now be classified 
according to their parity by solving separately 
$\det[D^{-1}_{nn'}({\bf k},\Omega)]^+
_{N^+ \times N^+}=0$ and $\det[D^{-1}_{nn'}({\bf k},\Omega)]^-
_{N^- \times N^-}=0$, respectively. 
This leads to $N^+ + 1$ even modes and $N^-$ odd ones. 
Moreover, it is possible to see from Eq.~(\ref{Dnm-2}) that
$[D^{-1}_{nn'}({\bf k},\Omega)]^-_{N^- \times N^-}$ is independent of 
$I({\bf k},\Omega)$. This means that correlations between local moment and 
band configurations do not influence 
spin modes of odd parity. As a consequence these modes are dispersionless, \cite{note4}
as we see explicitly below.  
More interesting are the even modes for which 
correlation effects due to the coupling to itinerant carriers
show up.

In the following we apply the above formulation to the cases of dilute and 
moderate Mn doping in the limit of vanishing spin-orbit coupling
($\gamma \rightarrow 0$). We then (Sec.~\ref{soc}) comment on how these
results are altered by a finite $\gamma$.

\subsection{Dilute Mn doping}
\label{ldk}

For illustration we start discussing the limiting case of dilute Mn doping 
or, equivalently, narrow quantum wells. This corresponds to very few Mn ions 
across the 
quantum well leading in our approach to a low-dimensional kernel 
$[D^{-1}_{n n'}({\bf k},\Omega)]_{N \times N}$ (Eq.~(\ref{Dnm-2})) with 
$N$ of order one. In this situation we can easily approach the problem 
analytically.
We choose for simplicity $N^+=N^-=1$ ({\it i.e.} $N=2$). 
In this case the dispersion 
relations of even and odd spin modes (corresponding to $\omega_0(z)$ and 
$\omega_1(z)$, respectively; see Eq.~(\ref{wn})) are obtained by solving 
\bea
\det[D^{-1}_{n n'}({\bf k},\Omega)]^+&=&[1+I({\bf k},\Omega)]x_{\rm s}|\Delta|-\Omega=0, 
\label{deven} \\
\det[D^{-1}_{n n'}({\bf k},\Omega)]^-&=&x_{\rm s}|\Delta|/2-\Omega=0 \, . 
\label{dodd}
\eea
Eq.~(\ref{dodd}) leads to a single 
odd mode with flat dispersion $\Omega^-=x_{\rm s}|\Delta|/2$.
For the calculation of the even modes we limit ourselves for now to the case of 
ferromagnetic coupling ($\Delta < 0$) at $T=0$; the difference between 
ferromagnetic and antiferromagnetic cases is commented on later 
in Sec.~\ref{CC}.
We solve Eq.~(\ref{deven}) 
for long and short wavelengths ({\it i.e.} long and short range correlations), 
using the expansions (\ref{int2}) and (\ref{int3}) up to first order in 
$\epsilon_{\bf k}$ and $1/\epsilon_{\bf k}$, respectively.
This leads to two solutions: one soft mode 
$\Omega^+_{\rm soft} < x_{\rm s}|\Delta|$, and one hard mode
$\Omega^+_{\rm stiff} \sim |\Delta|$, where typically $\Omega^+_{\rm soft} \ll 
\Omega^+_{\rm stiff}$.
For half-metallic carriers 
($|p|=1$, $\epsilon_{\rm s} \le |\Delta|$, see Appendix \ref{AppSP}) we obtain
for the small-and large-momentum limit
\bea
\Omega^+_{\rm soft}({\bf k})&=&\frac {x_{\rm s}}{1+x_{\rm s}}
\left(1-\frac{\epsilon_{\rm s}}{|\Delta|}\right) \epsilon_{\bf k}
+O(\epsilon_{\bf k}^2) 
\label{acem1dtp}\\
&=&x_{\rm s} |\Delta| \left(1- \frac{|\Delta|}{\epsilon_{\bf k}}\right)
+O(1/\epsilon_{\bf k}^2),
\label{acem1dtp2}\\
\Omega^+_{\rm stiff}({\bf k})&=&(1+x_{\rm s}) |\Delta| + 
\frac {1}{1+x_{\rm s}} 
\left(1+\frac{\epsilon_{\rm s}}{x_{\rm s}|\Delta|}\right) \epsilon_{\bf k}+O(\epsilon_{\bf k}^2), \nn \\
\label{opem1dtp}
\eea
respectively.
Correspondingly, for partly polarized carriers ($|p|<1$, 
$\epsilon_{\rm s} > |\Delta|$, using the results summarized in Appendix \ref{AppSP}) 
we find 
\bea
\Omega^+_{\rm soft}({\bf k})&=&0+O(\epsilon_{\bf k}^2)
\label{acem1dpp}\\
&=& x_{\rm s} |\Delta| \left(1+ \left( 1-\frac{2 \epsilon_{\rm s}}{|\Delta|}\right) \frac{|\Delta|}{\epsilon_{\bf k}} \right)+O(1/\epsilon_{\bf k}^2),
\label{acem1dpp2}\\
\Omega^+_{\rm stiff}({\bf k})&=&(1+x_{\rm s}) |\Delta| +
\frac{1}{x_{\rm s}}
\left(\frac{2 \epsilon_{\rm s}}{|\Delta|}-1 \right) \epsilon_{\bf k}+O(\epsilon_{\bf k}^2),
\label{opem1dpp}
\eea
for small and large momenta, respectively.

The branch $\Omega^+_{\rm soft}$ 
corresponds to a gapless Goldstone-mode reflecting the spontaneous 
breaking of rotational symmetry for the Mn spins subsystem, as 
expected generically in ferromagnets and found in bulk magnetic semiconductors. \cite{KMd-bulk} 
At long wavelengths 
($\epsilon_{\bf k} \rightarrow 0$) the dispersion 
in bulk isotropic ferromagnets is proportional to the spin stiffness 
$\rho$ divided by the magnetization 
${\cal M}$ ({\it i.e.} $\Omega=(\rho/{\cal M})k^2$). Similarly, in the 
adiabatic limit 
$\epsilon_{\rm s} \ll |\Delta|$, our 
long wavelength result for quantum wells (\ref{acem1dtp}) shows a spin 
stiffness due only to the 
increase in kinetic energy of the fully spin-polarized band when the spin 
orientation is spatial dependent, $\rho=\hbar^2 n^\da/(4 m^*)$. The 
magnetization, with parallel contributions from Mn ions and itinerant carriers
coupled ferromagnetically, reads ${\cal M}=S N_{\rm Mn}d+n^\da/2=S N_{\rm Mn}d(1+x_{\rm s})$. 
However, unlike bulk systems the spin stiffness vanishes 
as $\epsilon_{\rm s} 
\rightarrow |\Delta|$, Eq.~(\ref{acem1dtp}), and stays equal to zero for 
$\epsilon_{\rm s} > |\Delta|$, Eq.~(\ref{acem1dpp}).  This unusual feature 
should lead to some non-standard phenomenology in these ferromagnets, for example
in the physics that controls domain wall widths and finite-temperature magnetization
suppression.  
For short wavelengths ($\epsilon_{\bf k} \rightarrow \infty$), 
Eqs.~(\ref{acem1dtp2}) and (\ref{acem1dpp2}), the excitation energy 
$\Omega^+_{\rm soft}$ tends to a mean-field value $x_{\rm s} |\Delta|$, 
corresponding to the magnetic-ion spin splitting.

The branch of stiff excitations $\Omega^+_{\rm stiff}$, Eqs.~(\ref{opem1dtp}) and 
(\ref{opem1dpp}), is primarily band-like in character and 
is centered around the much larger energy 
scale of the itinerant carrier mean-field spin splitting $|\Delta|$.

\subsection{Moderate Mn doping}
\label{hdk}

\begin{figure} [h]
\begin{center}
\includegraphics[width=7cm,angle=0]{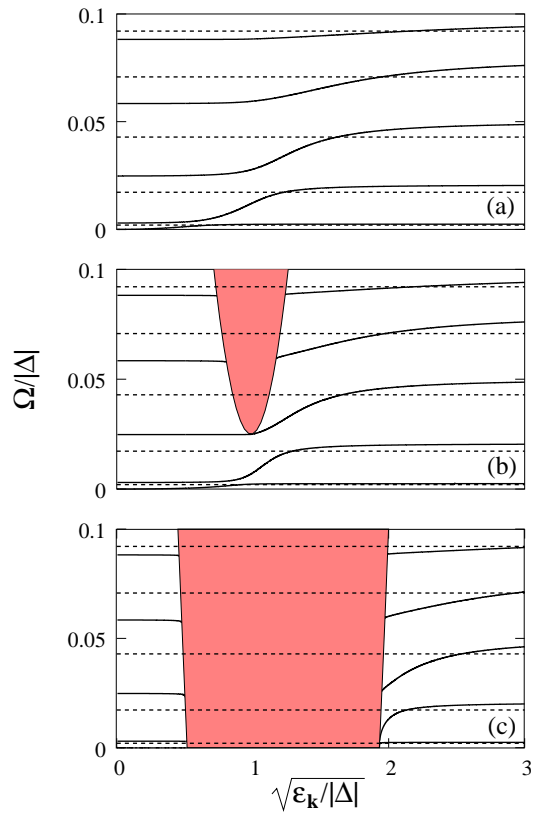}
\end{center}
\caption{
Dispersion of low-energy spin excitations in a single-subband DMS quantum 
well with 
ferromagnetic coupling ($\Delta < 0$) and dimensional cut-off $N^+=N^-=5$ 
(see text). Panels (a) and (b) correspond to half-metallic ({\it i.e.} fully polarized) carriers
($\epsilon_{\rm s}/|\Delta|= 0.9$ and 0.975, respectively). Panel (c) depicts
results for partly polarized carriers ($\epsilon_{\rm s}/|\Delta|= 1.5$).  
The even modes $\Omega^+_{(l)}$ are denoted by solid lines and the 
dispersionless odd modes $\Omega_{(l)}^-$ by dashed lines. The spatial profiles
of the even and odd modes are illustrated in  
Fig.~\ref{fig-5}(a)-(e) and Fig.~\ref{fig-7}, respectively.  The integer labels order the modes 
by increasing frequency. The shaded zones indicate the Stoner continuum.  
}
\label{fig-3}
\end{figure}

\begin{figure} [h]
\begin{center}
\includegraphics[width=7cm,angle=0]{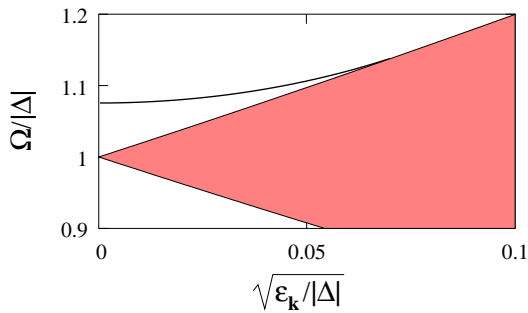}
\end{center}
\caption{
Dispersion of high-energy spin excitations in a single-subband DMS quantum 
well with ferromagnetic coupling ($\Delta < 0$) and dimensional cut-off 
$N^+=5$ (see text) for half-metallic ({\it i.e.} fully polarized) carriers
with $\epsilon_{\rm s}/|\Delta|= 0.9$. Similar features appear for partly 
polarized carriers ($\epsilon_{\rm s}/|\Delta| > 1$).
The stiff branch $\Omega^+_{\rm stiff}$ (solid lines) corresponds to the 
mode $\omega^+_{\rm stiff}$ (Fig.~\ref{fig-5}(f)). The Stoner continuum is 
represented by the shaded zone.   Ferromagnetic interactions with the local
moments peel a collective mode off the particle-hole continuum.
}
\label{fig-4}
\end{figure}

We now switch to the case of higher dimensions ($N^+, N^- > 1$), 
which corresponds to several Mn ions across the quantum well. 
As in the previous 
Sec.~\ref{ldk} dedicated to dilute Mn doping, we consider Mn spins 
coupled ferromagnetically to the itinerant carriers ($\Delta < 0$) at $T=0$. 
Comments regarding the 
antiferromagnetic coupling case will be introduced later in Sec.~\ref{CC}. 
As an example we choose $N^+ = N^- = 5$ with
a relative spin density $x_{\rm s}=0.05$. The 
dispersions corresponding to even and odd modes are obtained 
numerically by solving 
$\det[D^{-1}_{nn'}({\bf k},\Omega)]^+_{5 \times 5}=0$ and 
$\det[D^{-1}_{nn'}({\bf k},\Omega)]^-_{5 \times 5}=0$, respectively. 
This leads to a set of six even modes,
five relatively soft ($\Omega^+_{(l)}$) and one 
hard ($\Omega^+_{\rm stiff}$), and five odd modes 
($\Omega_{(l)}^-$). The index $l$ ($1\le l \le 5$) orders the modes from the 
bottom ($l=1$) to the top ($l=5$) of the spectrum in each case.
Our results for $\Omega^+_{(l)}$ (solid lines) and $\Omega_{(l)}^-$ 
(dashed lines) are 
summarized in Fig.~\ref{fig-3}(a)-(c) for three characteristic ratios
$\epsilon_{\rm s}/|\Delta|=$ 0.9, 0.975, and 1.5, respectively. 
Panels (a) and (b) correspond to half-metallic carriers while panel (c) 
depicts results for the case of partly polarized carriers.
Related results for $\Omega^+_{\rm stiff}$ are shown in Fig.~\ref{fig-4}. In all
plots the shaded zones represents the Stoner continuum. 
For each case, the normalized excitation energies $\Omega/|\Delta|$ are 
plotted as a function of the normalized $k$-vector
$\sqrt{\epsilon_{\bf k}/|\Delta|}= A (k/k_{\rm D})$, where 
$A=\sqrt{\epsilon_{{\bf k}_{\rm D}}/|\Delta|}$. For typical DMS quantum well
systems we estimate that  
$A > 10$. Hence, the results shown in Figs.~\ref{fig-3} and \ref{fig-4} correspond to 
$k/k_{\rm D} \ll 1$, the regime well below the Debye cut-off in which the continuum
approximation is most reliable.  
 
We start by discussing the properties of the relatively soft even modes 
$\Omega^+_{(l)}$ depicted in Fig.~\ref{fig-3} by solid lines. Panel (a)
corresponding to $\epsilon_{\rm s}/|\Delta|=$ 0.9 is representative of 
results for half-metallic cases with $\epsilon_{\rm s}/|\Delta| \ll 1$.
There we find a set of modes at ${\bf k}=0$ that are distributed in the small energy window 
$0 \le \Omega < 2 x_{\rm s} |\Delta|$ and have $k^2$ dispersion at finite
wavevectors, corresponding to a finite spin-stiffness. The 
upper limit in this spectrum is twice the mean-field value predicted in 
the case of bulk systems. \cite{KMd-bulk} This is 
due to the fact that the carrier spin density is modulated by 
$\chi(z)=\sqrt{2/d} \sin(\pi z/d)$ 
across the quantum well (lower density close to the border $z=0,d$ and higher
density close to the center $z=d/2$).
Spin modes with large relative amplitude 
near the center of the quantum well see a carrier spin density which is 
effectively higher than in the uniform spin-density bulk case.

\begin{figure} [h]
\begin{center}
\includegraphics[width=8cm,angle=0]{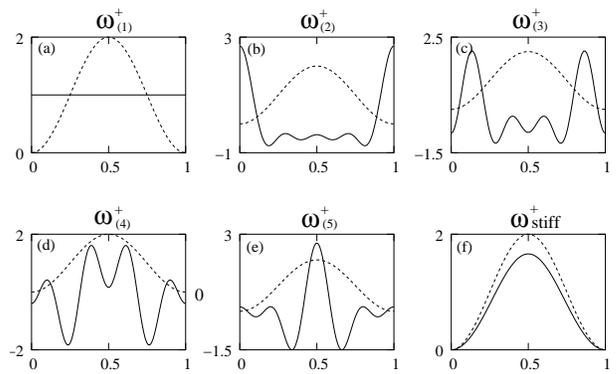}
\end{center}
\caption{
Spatial profiles across a single-subband DMS quantum well for even parity 
collective modes (solid lines).  The well has width $d$ ($0 \le z/d \le 1$), 
ferromagnetic coupling ($\Delta < 0$), and the dimensional cut-off $N^+=5$ 
(see text). The carrier density $\chi^2(z)$ (dashed line) is 
shown for comparison.  Note that the Goldstone mode, panel (a), is constant in space 
corresponding to uniform spin rotation.  The other modes tend to have higher 
energy when they have higher weight toward the middle of the quantum well, where
exchange interactions are stronger.
}
\label{fig-5}
\end{figure}

\begin{figure} [h]
\begin{center}
\includegraphics[width=5cm,angle=-90]{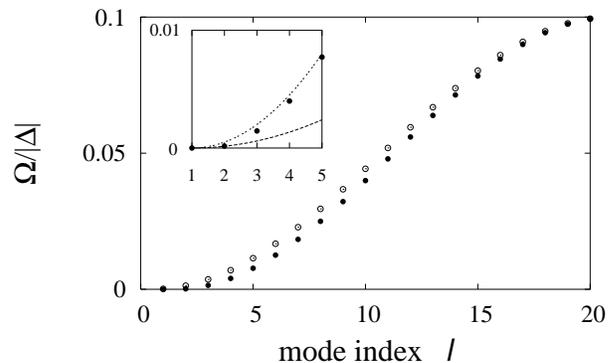}
\end{center}
\caption{
Dispersion of low-energy spin excitations in a single-subband DMS quantum 
well for in-plane momentum ${\bf k}=0$, ferromagnetic coupling 
($\Delta < 0$), and dimensional cut-off $N^+=N^-=20$, as a function of 
the mode 
index $l$ (see text). The full circles depict the excitation energies
$\Omega^+_{(l)}({\bf k}=0)$ of even modes $\omega^+_{(l)}$, while the empty 
circles correspond to $\Omega^-_{(l)}({\bf k}=0)$ for 
odd modes $\omega^-_{(l)}$. The inset corresponds to two extreme quadratic fittings (dashed and 
dotted curves, see text) to the dispersion of even modes (full circle) for small $l$. 
The curves differ by a prefactor of order 3.5.
}
\label{fig-6}
\end{figure}

The first branch $\Omega^+_{(1)}$ corresponds to a gapless 
Goldstone-mode $\omega^+_{(1)}(z)=\omega_0(z)$, Fig.~\ref{fig-5}(a),
similar to the one discussed above for the dilute Mn doping case, 
in Sec. \ref{ldk}. 
For the $l > 1$ modes, the spatial structure $\omega^+_{(l)}(z)$ is not constant within the 
quantum well (see Figs.~\ref{fig-5}(b)-(e)) and the energies $\Omega^+_{(l)}({\bf k})$ 
therefore approach a finite value as $|{\bf k}| \to 0$.
The dependence of the excitation energies $\Omega^+_{(l)}({\bf k}=0)$ 
on $l$ is not obvious, since the gap is not simply related to the 
effective transverse 
momentum $k_z^{+(l)}=\int_0^d {\rm d}z~\omega^+_{(l)} 
\partial \omega^+_{(l)}/\partial z$ 
associated to each mode. Instead,  
the local excitation density $\omega^{+ 2}_{(l)}(z)$ and its correlation 
with the carrier density, proportional to $\chi^2(z)$, is more relevant. (See Fig. \ref{fig-5} for a 
comparison). We illustrate the $l$-dependence of 
$\Omega^+_{(l)}({\bf k}=0)$ in Fig.~\ref{fig-6} for $N^+=20$ (full circles).  We observe a 
spin-mode accumulation close to the boundaries of the spectrum that is not evident 
for small $N^+$. Furthermore, the inset in Fig.~\ref{fig-6} shows that the dispersion 
does \emph{not} appear to be quadratic for small $l$. Since $\Omega^+_{(1)}({\bf k}=0)=0$, 
a quadratic fit has a single free parameter. The dashed (dotted) line 
corresponds to a quadratic fit between the $l=1$-mode and the $l=2$- ($l=5$) one. The curves
differ by a (relatively large) factor of order 3.5.  

As $\epsilon_{\rm s}/|\Delta|$ increases (Fig.~\ref{fig-3}(b)) and the Stoner 
continuum meets the different $\Omega^+_{(l)}$ branches, the 
corresponding spin stiffness drops to zero. 
For $\epsilon_{\rm s}/|\Delta| \ge 1$ (partly spin-polarized carriers, 
Fig.~\ref{fig-3}(c)) all branches are nearly dispersionless until they enter the 
particle-hole continuum; the softness of these excitations is certain to have an impact
on the magnetic properties of these ferromagnets.  
The spectral density shows that the relatively soft modes 
survive in the midst of the Stoner continuum, but the
stiff mode does not as discussed below. 

Regarding Mn doping and kernel dimensionality, $N^+$ increases with higher
magnetic-ion density and quantum well width. In this situation,
new, relatively soft, modes arise from the top of the spectrum squeezing 
the rest to the bottom in order to satisfy 
$0 \le \Omega^+_{(l)} < 2 x_{\rm s} |\Delta|$. This happens because 
increasing the dimension of the kernel admits the presence of higher order Fourier components
in our expansion and allows lower energy modes to aquire a larger relative amplitude at the borders 
of the quantum well, where the free-carrier density is reduced and
the Mn spin splitting is consequentially weaker. The same consideration applies to 
the odd mode (see below) behavior as a function of $N^-$.
 
In addition to the relatively soft modes we find one even, stiff branch 
$\Omega^+_{\rm stiff}$ (Fig. \ref{fig-4}) lying above the Stoner continuum. 
Its properties are similar to those discussed for the stiff mode in the 
dilute Mn doping case, Sec.~\ref{ldk}. The mode is restricted to 
relatively 
large wavelengths if compared with the case of soft modes. This is 
due to the proximity of 
the Stoner continuum and their strong interaction. The corresponding 
excitation 
profile $\omega_{\rm stiff}^+(z)$ is shown in Fig.~\ref{fig-5}(f) (solid line). 
Its similarity to the carrier density profile $\chi^2(z)$ (dashed line) demonstrates that 
this mode is primarily associated with the itinerant-carrier subsystem dynamics. 

\begin{figure} [h]
\begin{center}
\includegraphics[width=8cm,angle=0]{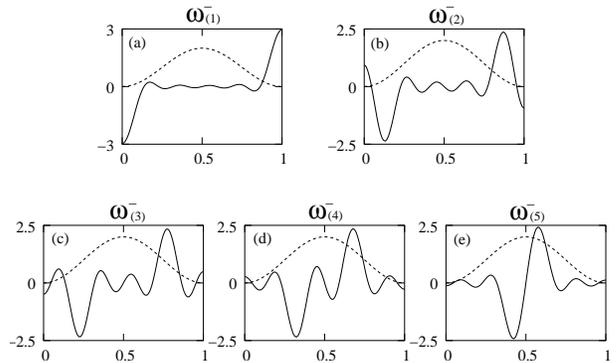}
\end{center}
\caption{
Spatial profiles for spin excitation modes with odd parity (solid lines) across a single-subband 
DMS quantum well of width $d$ ($0 \le z/d \le 1$) with 
ferromagnetic coupling ($\Delta < 0$) and dimensional cut-off $N^-=5$ 
(see text). The transverse carriers density $\chi^2(z)$ (dashed line) is 
shown for comparison.
}
\label{fig-7}
\end{figure}

We continue discussing briefly the properties of the 
dispersionless odd modes $\Omega_{(l)}^-$ depicted in Fig.~\ref{fig-3} (dashed lines). 
As pointed out above, the 
flat dispersions have their origin in the absence of correlation
effects related to spin reorientations. Odd mode fluctuations of the 
local moment give rise to effective fields that are averaged to zero
by $\chi^{2}(z)$ and are consequently not correlated with carrier spin
fluctuations.  This makes the odd modes transparent to the
Stoner excitations (they do not interact with the particle-hole 
continuum, unlike the even modes) and independent of the ratio 
$\epsilon_{\rm s}/|\Delta|$ (that is, of the carrier spin polarization).
Moreover, all modes are lodged within the (low) energy window 
$0 < \Omega < 2 x_{\rm s} |\Delta|$ and present a finite gap whose magnitude 
depends on the particular excitation profile $\omega^-_{(l)}(z)$ 
(Fig.~\ref{fig-7}). This behavior is similar to that found for the relatively soft even 
modes except that there is no Goldstone mode in the odd spectrum; 
the higher the weight at the border of the quantum well, the 
lower the gap (see {\it e.g.} $\omega^-_{(1)}(z)$, Fig.~\ref{fig-7}(e), which corresponds to 
$\Omega_{(1)}^-$ in Fig.~\ref{fig-3}). As in the even case, the dependence of 
$\Omega^-_{(l)}$ 
on $l$ is nontrivial. See Fig.~\ref{fig-6} (empty circles) 
for an illustration of this dependence in the case $N^-=20$.

We further note that, as can be seen from Eq.~(\ref{Dnm-2}), 
in the limit of a large number of vertical modes ($N^+,N^- \gg 1$) 
it holds $\Omega_{(l)}^- \approx \Omega^+_{(l+1)}({\bf k}=0)$, provided that 
$x_{\rm s} \ll 1$.  The even and odd mode pairs are then nearly degenerate 
even and odd combinations of excitations at opposite edges of the 
quantum well.\cite{note5}   

Our analysis holds for symmetric quantum wells only.
We comment shortly on how an asymmetry of the quantum-well potential will
affect or results.
On the one hand, a gapless Goldstone mode as found for 
symmetric quantum wells when neglecting spin-orbit coupling (see also 
Sec. \ref{soc}) still exists.
On the other hand, the spin-wave modes will have no definite parity anymore,
and the above classification into even (dispersive) and odd (nondispersive)
modes no longer holds.
The particular excitation profiles (Figs. \ref{fig-5} 
and \ref{fig-7}) and energies (Fig. \ref{fig-6}) will depend on the 
local-carrier density $\chi^{2}(z)$.

\subsection{Effect of spin-orbit coupling}
\label{soc}

The presence of spin-orbit coupling described by the Dresselhaus Hamiltonian 
$H_{\rm D}$, Eq.~(\ref{HD}), introduces easy-axis magnetic anisotropy which is,
of course, necessary for long-range magnetic order in a quantum well.
When the anisotropy, which explicitly breaks rotational invariance for the magnetization 
orientation, is accounted for a finite energy-gap $\Omega_{\rm so}$ appears 
in the lowest lying collective mode branch and several of the 
lower-lying branches in Figs.~\ref{fig-3} and \ref{fig-6} are shifted 
to higher energies. We calculate $\Omega_{\rm so}$ for the 
lowest even mode of constant excitation profile $\omega_{(1)}^+$, Fig.~\ref{fig-5}(a). 
With this aim we follow the procedure of Sec.~\ref{DMnD} and calculate 
$\Omega_{(1)}^+$ for small spin-orbit coupling and ${\bf k}=0$. We find that 
\bea
\Omega_{\rm so}=\frac {\gamma^2 \langle k_z^2 \rangle^2 k_{\rm F}^2 n}
{S N_{\rm Mn} d~{\rm max} \{ |\Delta|,2\epsilon_{\rm F} \} },
\eea
where $\epsilon_{\rm F}=(\hbar^2/2 m^*) k_{\rm F}^2$ is the Fermi energy of 
the 2D electron gas paramagnetic state and $n=n^\ua+n^\da$.
This coincides with our previous result \cite{Koenig} obtained by using 
perturbation theory.

The above result and discussion holds for symmetric quantum wells.
In asymmetric quantum wells, spin-orbit coupling is not only described 
by the Dresselhaus terms, but has also a Rashba \cite{rashba} spin-orbit 
contribution.
For electrons in the conduction band, the Rashba Hamiltonian is linear in the 
in-plane momentum ${\bf k}$.
This will affect the magnetic anisotropy and the energy gap in the spectrum of 
collective spin excitations.
The interplay of these two types of spin-orbit coupling will complicate the
dispersion of the spin waves (Fig. \ref{fig-3}).
In particular, they will become anisotropic, analogous to the anisotropic 
transport properties discussed in Ref.~\onlinecite{schliemann}.
For heavy holes in the valence band the leading Rashba term is cubic in 
momentum.\cite{W00} 
This complicates the evaluation of the spin-wave dispersions even more.

\section{Comments and conclusion}
\label{CC}

We have developed a theory of collective spin excitations in DMS quantum wells 
by extending the approach that we used previously for bulk systems. \cite{KMd-bulk} 
The theory goes beyond mean-field and RKKY approaches and accounts for both
finite itinerant carrier spin-splitting and dynamical correlations. 
We applied this tool to the study of spin excitations in the ordered magnetic 
state at zero temperature.  As in the bulk case, we have recognized two 
different 
energy scales on which the spin excitation spectrum depends: one hard scale 
$|\Delta|$ principally related to the itinerant-carrier subsystem, and one soft 
scale $x_{\rm s}|\Delta|$ for the magnetic-ion spin excitations, 
where $x_{\rm s}|\Delta| \ll |\Delta|$. In addition, a continuum of Stoner 
excitations (corresponding to flipping a single spin in the itinerant-carrier 
band) also emerges from the theory.  Although most relevant to DMS ferromagnet 
properties in circumstances for which the  
magnetic moments have a high degree of spin alignment, this theory of the elementary excitation 
spectrum of the system sheds considerable light on the nature of the magnetic state
and on the physics that controls the critical temperature of the system.

The excitation spectrum of this magnetic system is quite unusual because of its
ambiguous dimensionality.  A slab of magnetic ions is coupled by a 2D electron system that 
is frozen into a single growth direction electronic subband and cannot 
distort its $z$-dependence to accommodate magnetic fluctuations.  
We find that the excitation spectrum of this system has multiple 2D branches. The number of 
reasonably well-defined branches of excitations that have primarily local moment 
character is close to the width of the quantum well measured 
in units of the mean-separation between Mn ions, as expected by analogy with a
reference systems in which the local moments are placed on a lattice with 
the same volume per Mn and a finite number of layers.  On the other hand,
we find that there is only one 2D branch of collective excitations that 
have primarily electronic character.  When spin-orbit interactions are neglected,
the gapless Goldstone mode branch has quartic rather than quadratic dispersion, 
implying that the spin-stiffness vanishes, except when the carrier system is half-metallic.
Unless, that is, $\epsilon_{\rm s} < |\Delta|$
and the mean-field carrier spins are consequently fully spin-polarized.  This 
property arises somewhat accidentally from the particular features of effective 
interactions mediated by carriers in 2D parabolic bands and has partly been
noted for the RKKY limit ($\epsilon_{\rm s} \gg |\Delta|$) in previous 
work.\cite{KFACTWdSGBSSWD00}
If we had included carrier-carrier interactions in our theory, the spin-stiffness would not
vanish but, depending on the carrier density, might have a negative sign 
implying that the ferromagnetically ordered state is unstable.  When 
spin-orbit interactions are included, however, the collective excitation spectrum
has a small gap and negative dispersion in the lowest-lying collective mode, while unusual, does
not necessarily imply instability. Finally, we remark that the excitation spectrum includes
a large number of even and odd weakly dispersive or non-dispersive branches
in which fluctuations in the local moments are concentrated in particular 
parts of the quantum well and do not couple strongly to fluctuations in
the band-electron spin-orientation.  In these modes, the excitation energy is determined 
primarily by the local strength of the mean-field interaction between the 
Mn moments and the band electrons, which becomes small because of carrier quantum-size 
effects toward the edges of the quantum well.     

The results summarized above for the collective excitation spectrum suggest
that thermodynamic properties, the temperature dependence of 
the magnetization for example, are likely to be quite unusual in DMS ferromagnets,
particularly when the carriers are in the conduction band where
spin-orbit interactions are rather weak.  
The influence of thermal fluctuations on the magnetization will be enhanced not
only by the reduced dimension \cite{MW66}, but also by the small and possibly
even negative spin stiffness mentioned above.  (Fluctuations in the Goldstone
mode branch have a relative importance that goes like $1/N$, where $N$ is
the effective number of Mn layers in the film that we have discussed previously.)
In the case of valence band 
DMS quantum well ferromagnets, strong magnetic anisotropy \cite{LeeThesis} should
lead to ferromagnetism that is essentially Ising in character.  The magnetization
should then be fairly constant over a wide interval of temperature before dropping
fairly rapidly to zero near the critical temperature.  In the conduction band 
case for which the model we have studied applies most directly, however, the gap
in the excitation spectrum will be quite small, much smaller than the mean-field
ferromagnetic critical temperature, as we have discussed earlier.\cite{Koenig} 
The true critical temperature is likely to be 
determined in large measure by long length scale fluctuations and to be 
substantially smaller than the mean-field temperature. Since the stiffness
of this system is very small, it will be significantly altered by spin-orbit interactions;
this part of the physics is something that we have not addressed here.  

Furthermore, the model
we have studied ignores disorder, which is likely to play an important role in adjudicating the 
way in which these subtle competitions are resolved.  We believe that careful study of 
the magnetization and other properties of electron quantum well DMS ferromagnets 
at low temperatures will reveal a lot of subtle, unusual, and interesting physics.    

Finally we emphasize that the numerical results presented here are 
for the case of ferromagnetic interactions between the carriers and 
the local moments ($\Delta < 0$), which is expected to apply for 
n-doped semiconductors.  Since we have taken the local-moment spins to 
point down in their ground state, this means that the up spins are
the minority spins and the down spins are the majority spins in the 
ferromagnetic case while their roles are interchanged in the antiferromagnetic
case.  It follows that the only change in Eq.~\ref{D} when $\Delta$ changes
sign is that $i\nu_n$ changes sign.  This change has a number of consequences that 
are fairly subtle when $x_s$ is small, but can in principle be more consequential.
The most important differences are that the collective mode with dominant 
carrier character which appears above the Stoner particle-hole continuum in
the case of ferromagnetic interactions, lies below the Stoner continuum in the case
of antiferromagnetic interactions.  In addition $1/(1+x_s)$ factors which appear 
in expressions for the collective mode energies, factors that express the band 
electron contribution to the total spin density of the system, are replaced by 
$1/(1-x_s)$ factors in the antiferromagnetic case.  These factors are not present 
in a RKKY description of the carrier mediated interactions.

\acknowledgments

We thank C. Balseiro, T. Jungwirth, Byounghak Lee, and U. Z\"ulicke for 
helpful discussions.
This work was supported by the Deutsche Forschungsgemeinschaft via the 
Emmy-Noether program and the Center for Functional Nanostructures, by the Research Training Network 
{\it Spintronics}, by the National Science Foundation under grant DMR 0210383, and by the Welch 
Foundation.

\appendix

\section{Mean-field itinerant carrier spin polarization}
\label{AppSP}

The 2D mean-field itinerant carrier spin density for spin $\sigma$ and 
subband spin-splitting $\Delta= J_{\rm ex} N_{\rm Mn} S$ in the presence of 
weak Dresselhaus spin-orbit coupling at $B=0$ is given 
by $n^\sigma=\int {\rm d}^2k/(2 \pi)^2 f(\epsilon_{\bf k}^\sigma)$, where  
$\epsilon_{\bf k}^\sigma=\epsilon_{\bf k}-(\sigma/2)\Delta+\sigma \epsilon_{\bf k}^{\rm so}-\mu'$, 
$\epsilon_{\bf k}=\hbar^2 k^2/(2 m^*)$, $\epsilon_{\bf k}^{\rm so}=-\gamma^2 \langle k_z^2 \rangle^2 k^2/\Delta$, $\mu'$ is the effective chemical potential of the 2D carrier gas, 
and $f(\epsilon)$ is the Fermi distribution. For zero temperature we find
\bea
n^\sigma=\frac {\mu'+(\sigma/2)\Delta}{4 \pi(\hbar^2/2m^*-\sigma \gamma^2 \langle k_z^2 \rangle^2/\Delta)}~
\theta(\mu'+(\sigma/2)\Delta),
\label{nspin}
\eea
where we have used $f(\epsilon)_{T=0}=\theta(-\epsilon)$.
The difference between up and down contributions determines the 
net carrier spin density defined as 
$p=(n^\ua-n^\da)/(n^\ua+n^\da)$. Denoting the Fermi energy of the 
majority-spin carrier band by $\epsilon_{\rm s}=\mu'+|\Delta|/2$, we see 
from Eq.~(\ref{nspin}) that the carrier system is fully spin-polarized or {\it half-metallic} 
({\it i.e.} $|p|=1$) when $\epsilon_{\rm s} \le |\Delta|$. On the other hand, partly 
spin-polarized carriers ({\it i.e.} $|p|<1$) correspond to $\epsilon_{\rm s} > 
|\Delta|$. 

\section{Correction due to correlation effects}
\label{AppCC}

Correlation effects due to the response of quantum well carriers to Mn 
spin reorientations are taken into account in the kernel 
$D^{-1}({\bf k},z,z',\Omega)$ by the second term of Eq.~(\ref{D}). 
This is manifested by the momentum and energy dependence of the integral
$I({\bf k},\Omega)$, Eq.~(\ref{IK}).  
In the absence of magnetic field and spin-orbit coupling we find for zero 
temperature (see Appendix~\ref{AppSP} for definitions)
\begin{widetext}
\bea
\int \frac {{\rm d}^2q}{(2 \pi)^2}~ 
\frac {[f(\epsilon_{\bf q}^\da)
-f(\epsilon_{{\bf q}+{\bf k}}^\ua)]_{T=0}}{\Omega + \epsilon_{\bf q}^\da - 
\epsilon_{{\bf q}+{\bf k}}^\ua}&=&
\frac{m^*}{4 \pi \hbar^2 \epsilon_{\bf k}}\left[
\theta(\mu'-\Delta/2)~(\Omega+\Delta-\epsilon_{\bf k}) 
\left(1-\sqrt{1-\displaystyle \frac{4 (\mu'-\Delta/2) \epsilon_{\bf k}}
{(\Omega+\Delta-\epsilon_{\bf k})^2}}~\right) \right. \nn \\
&-& \left. \theta(\mu'+\Delta/2)~(\Omega+\Delta+\epsilon_{\bf k}) 
\left(1-\sqrt{1-\displaystyle \frac{4 (\mu'+\Delta/2) \epsilon_{\bf k}}
{(\Omega+\Delta+\epsilon_{\bf k})^2}}~\right) \right] 
\label{int} \\
\nn \\
&=& \frac{n^\da-n^\ua}{\Omega+\Delta} + \frac{n^\da~(\mu'+\Delta/2+\Omega)
-n^\ua~(\mu'-\Delta/2-\Omega)}{(\Omega+\Delta)^3}~\epsilon_{\bf k} 
+ O(\epsilon_{\bf k}^2)
\label{int2} \\
\nn \\
&=&-(n^\ua+n^\da)/\epsilon_{\bf k} 
+ O(1/\epsilon_{\bf k}^2), 
\label{int3}
\eea
\end{widetext}
where the integral has to be considered as a Cauchy principal value. 
Eqs.~(\ref{int2}) and (\ref{int3}) correspond to first order expansions in $\epsilon_{\bf k}$ 
and $1/\epsilon_{\bf k}$, respectively.
The prefactors that express the $z$-dependence in the second term of 
Eq.~(\ref{D}) contain information on the quantum well geometry 
and lead to $M_2$ in Eq.~(\ref{Dnm-2}).


\begin{thebibliography}{99}
\bibitem{note10} Another related consequence is that the carrier-band and Mn 
mean-field spin splitting typically belong to different energy scales in DMSs.
\bibitem{note7} Starting with a description where the Mn distribution is 
treated as a continuum points out the difference between DMS and spin 
glass systems, where dilute Mn local spins are randomly distributed in a 
metallic host. In both cases the interaction between local spins mediated 
by the itinerant carriers is oscillatory, and ferromagnetic for distances $L$
shorter that the carriers Fermi wavelength $\lambda_{\rm F}$. However,
in the spin glass case it is $\lambda_{\rm F} < L$, while for DMSs we have
$\lambda_{\rm F} > L$. Hence, each Mn ion in a DMS interacts ferromagnetically 
with several of its neighbors. The continuum Mn model will 
fail for very dilute Mn impurities and in case the exchange interaction 
between Mn and carrier spins is too strong.\cite{A80}
\bibitem{note9} For simplicity we neglect here higher order (cubic) terms in $k_x$ and
$k_y$. However, the approximation is accurate for single-subband quantum wells.
\bibitem{note5} In the continuum approximation that we employ, the same result can be obtained even
for small numbers of vertical modes if the cut-offs for the even and odd modes are chosen 
asymmetrically so that $N^- = N^+ -1$.
\end{thebibliography}

\end{document}